\documentclass[times]{cpeauth}
\usepackage[dvips,colorlinks,bookmarksopen,bookmarksnumbered,citecolor=red,urlcolor=red]{hyperref}
\usepackage[nolist]{acronym}
\usepackage{mathptmx}
\usepackage{moreverb}
\usepackage{color}
\usepackage{graphicx}
\usepackage{listings}
\usepackage{array}
\usepackage{times}
\usepackage{rotating}
 \usepackage[font={small}]{caption, subfig}

\newcommand\BibTeX{{\rmfamily B\kern-.05em \textsc{i\kern-.025em b}\kern-.08em
T\kern-.1667em\lower.7ex\hbox{E}\kern-.125emX}}

\newcommand{\bq}{\begin{equation}}
\newcommand{\eq}{\end{equation}}

\newcommand{\flops}{\mbox{flops}}
\newcommand{\updateX}{\mbox{update}}
\newcommand{\updates}{\mbox{updates}}
\newcommand{\UP}{\mbox{UP}}
\newcommand{\US}{\mbox{\UP/\second}}
\newcommand{\GUS}{\mbox{G\UP/\second}}

\newcommand{\second}{\mbox{s}}

\newcommand{\GBS}{\mbox{GB/s}}

\newcommand{\GHZ}{\mbox{GHz}}
\newcommand{\GCS}{\mbox{G\cycles/\second}}

\newcommand{\bytes}{\mbox{bytes}}

\newcommand{\BYTE}{\mbox{B}}
\newcommand{\BYTES}{\mbox{B}}
\newcommand{\bit}{\mbox{b}}

\newcommand{\GB}{\mbox{GB}}

\newcommand{\MB}{\mbox{MB}}

\newcommand{\cycle}{\mbox{cy}}
\newcommand{\cycles}{\mbox{cy}}

\newcommand{\eos}{~.}

\newcommand{\doi}[1]{\href{http://dx.doi.org/#1}{DOI:#1}}

\newcommand{\ecmm}{ECM model}

\newcommand{\construction}[1]{}

\newcommand{\olsep}{\|}
\newcommand{\nolsep}{|}
\newcommand{\ecmspace}{\,}
\newcommand{\ecm}[6]{\mbox{$\left\{{#1}\ecmspace\olsep\ecmspace {#2}\ecmspace\nolsep\ecmspace {#3}\ecmspace\nolsep\ecmspace {#4}\ecmspace\nolsep\ecmspace {#5}\right\}\ecmspace{#6}$}}
\newcommand{\epsep}{\rceil}
\newcommand{\ecmp}[5]{\mbox{$\left\{{#1}\ecmspace\epsep\ecmspace {#2}\ecmspace\epsep\ecmspace {#3}\ecmspace\epsep\ecmspace {#4}\right\}\ecmspace{#5}$}}

\usepackage{sidecap}

\newcommand{\ecmknc}[5]{\mbox{$\left\{{#1}\ecmspace\olsep\ecmspace {#2}\ecmspace\nolsep\ecmspace {#3}\ecmspace\nolsep\ecmspace {#4}\right\}\ecmspace{#5}$}}
\newcommand{\ecmpknc}[4]{\mbox{$\left\{{#1}\ecmspace\epsep\ecmspace {#2}\ecmspace\epsep\ecmspace {#3}\right\}\ecmspace{#4}$}}
\newcommand{\ecmshort}[5]{\mbox{$\left\{{#1}\ecmspace\olsep\ecmspace {#2}\ecmspace\nolsep\ecmspace {#3}\ecmspace\nolsep\ecmspace {#4}\ecmspace\nolsep\ecmspace {#5}\right\}$}}
\newcommand{\ecmpshort}[4]{\mbox{$\left\{{#1}\ecmspace\epsep\ecmspace {#2}\ecmspace\epsep\ecmspace {#3}\ecmspace\epsep\ecmspace {#4}\right\}$}}

\def\volumeyear{2016}

\begin{document}

\begin{acronym}[IMCI]
    \acro{AGU}{address generation unit}
    \acro{AVX}{advanced vector extensions}
    \acro{CL}{cache line}
    \acro{CoD}{cluster-on-die}
    \acro{DP}{double precision}
    \acro{ECM}{execution-cache-memory}
    \acro{FMA}{fused multiply-add}
    \acro{IMCI}{initial many core instructions}
    \acro{LLC}{last-level cache}
    \acro{NUMA}{non-uniform memory access}
    \acro{SIMD}{single instruction multiple data}
    \acro{SP}{single precision}
    \acro{UFS}{uncore frequency scaling}
\end{acronym}

\runningheads{J. Hofmann et al.}{Performance analysis of the Kahan-enhanced scalar product}

\title{Performance analysis of the Kahan-enhanced scalar product on current multi- and manycore processors}

\author{J.~Hofmann\affil{1}\comma\corrauth{}\comma\footnotemark[2], D.~Fey\affil{1}, M.~Riedmann\affil{2}, J.~Eitzinger\affil{3}, G.~Hager\affil{3} and G.~Wellein\affil{3}}

\address{\centering
   \affilnum{1} Chair for Computer Architecture, University of Erlangen-Nuremberg, Erlangen, Germany\\
   \affilnum{2} AREVA GmbH, Erlangen, Germany\\
   \affilnum{3} Erlangen Regional Computing Center (RRZE), University of Erlangen-Nuremberg, Erlangen, Germany
}

\corraddr{Johannes Hofmann, Lehrstuhl f\"ur Rechnerarchitektur (Informatik 3), Martensstr. 3, 91058 Erlangen, Germany}
\footnotetext[2]{E-mail: johannes.hofmann@fau.de}

\begin{abstract}
%
  We investigate the performance characteristics of a numerically
  enhanced scalar product (dot) kernel loop that uses the Kahan
  algorithm to compensate for numerical errors, and describe efficient
  SIMD-vectorized implementations on recent multi- and manycore
  processors. Using low-level instruction analysis and the
  execution-cache-memory (ECM) performance model we pinpoint the
  relevant performance bottlenecks for single-core and thread-parallel
  execution, and predict performance and saturation behavior.  We show
  that the Kahan-enhanced scalar product comes at almost no additional
  cost compared to the naive (non-Kahan) scalar product if appropriate
  low-level optimizations, notably SIMD vectorization and unrolling,
  are applied. The ECM model is extended appropriately to accommodate
  not only modern Intel multicore chips but also the Intel Xeon Phi
  ``Knights Corner'' coprocessor and an IBM POWER8 CPU. This allows us
  to discuss the impact of processor features on the performance
  across four modern architectures that are relevant for high
  performance computing.

\end{abstract}


\maketitle

\section{Introduction and related work}

Accumulating finite-precision floating-point numbers in a scalar variable is
a common operation in computational science and engineering. The consequences
in terms of accuracy are inherent to the number representation and have been
well known and studied for a long time~\cite{Goldberg:1991}. There are a number
of summation algorithms that enhance accuracy while maintaining an acceptable
throughput~\cite{Linz:1970,Gregory:1972}, of which  Kahan~\cite{Kahan:1965}
is probably the most popular one. However, the topic is still subject to active
research \cite{Rump:2008,Zhu:2010,Demmel:2013,Dalton:2014}. A straightforward
solution to the inherent accuracy problems is arbitrary-precision floating
point arithmetic, which comes at a significant performance penalty.
Naive summation and arbitrary precision arithmetic are at opposite ends of
a broad spectrum of options, and balancing performance vs.\ accuracy is a key
concern when selecting a specific solution.

Naive summation, which simply adds each successive number in sequence
to an accumulator, requires appropriate unrolling for \ac{SIMD}
vectorization and pipelining. The necessary code transformations are
performed automatically by modern compilers, which results in optimal
in-core performance.  Such a code quickly saturates the memory
bandwidth of modern multi-core CPUs when the data is in memory.

This paper investigates implementations of the scalar product, a kernel which
is relevant in many numerical algorithms. Starting from an optimal naive
implementation it considers scalar and SIMD-vectorized versions of the Kahan
algorithm using various SIMD instruction set extensions on a range of multi-
and manycore processors from Intel and IBM.  Using an analytic performance
model we point out the conditions under which Kahan comes for free, and we
predict the single core performance in all memory hierarchy levels as well as
the scaling behavior across the cores of a chip. The present work is an extended version of
\cite{HFEHW15}, where we carried out the analysis for a range of older Intel Xeon
processors. Apart from new architectures we present a refined version
of the ECM performance model and add an additional optimization for the Intel
Haswell-EP CPU\@.

This paper is organized as follows. In Sect.~\ref{sec:testbed} we give an overview
of the hardware used for analysis and benchmarking. Section~\ref{sec:models}
introduces the execution-cache-memory (ECM) performance model, which is
used in Sect.~\ref{sec:impl} to describe different variants of the naive
and the Kahan scalar product. Section~\ref{sec:results} gives performance
results and validates the models. Section~\ref{sec:conc}
provides a conclusion and some comments on the possible extension of
our work.

\section{The ECM performance model}\label{sec:models}
The \ac{ECM} model \cite{Treibig:2009,hager:cpe,sthw15,HFEHW15} is an
analytic performance model that uses hardware architecture specifications and
few measurements as input. It estimates the number of CPU cycles
required to execute a number of iterations of a loop on a single core of a
multi- or many-core chip. The prediction comprises contributions from the
in-core execution time $T_\mathrm{core}$, i.e., the time spent executing
instructions in the core under the assumption that all data resides in the L1
cache, and the transfer time $T_\mathrm{data}$, i.e., the time spent
transferring data from its location in the cache/memory hierarchy to the L1
cache. As data transfers in the cache and memory hierarchy occur at \ac{CL}
granularity we choose the number of loop iterations $n_\mathrm{it}$ to correspond to one cache
line's ``worth of work.'' On Intel architectures, where \ac{CL}s are
64\,\BYTE\ long, we use  $n_\mathrm{it}=16$ for the \ac{SP} dot product because
sixteen \ac{SP} floating-point numbers (4\,\BYTE\ each) fit into one \ac{CL}. 
\ac{CL}s on the IBM POWER8 architectures are
128\,\BYTE, which leads to $n_\mathrm{it}=32$ for the \ac{SP} dot product.

Superscalar core designs house multiple execution units, for
loading and storing data, multiplication, division, addition, etc.
The in-core execution time $T_\mathrm{core}$ is determined by the unit that
takes the longest to execute the instructions allocated to it.  Other
constraints for the in-core execution time of a single core may apply, e.g., the
four micro-op per cycle retirement limit for Intel's Xeon cores and the eight
instruction per cycle retirement limit for IBM's POWER8 core.
The model differentiates between core cycles depending on whether data
transfers in the cache hierarchy can overlap with in-core execution time.
For instance, on Intel Xeons, core cycles in which data is moved between
the L1 cache and registers, e.g., cycles in which load and/or store instructions
are retired, prohibit the simultaneous transfers of data between the L1 and L2
cache; these ``non-overlapping'' cycles contribute to $T_\mathrm{nOL}$. Cycles
in which other instructions, such as arithmetic instructions, retire are
considered ``overlapping'' cycles and contribute to $T_\mathrm{OL}$. The
in-core runtime is the maximum of both: $T_\mathrm{core} = \max(T_\mathrm{OL},
T_\mathrm{nOL})$. Note that the non-overlapping quality of L1-register transfers
is specific to Intel CPUs. We will see later that the IBM POWER8 does not
have non-overlapping instructions.

For modeling the data transfers, latency effects are initially neglected, so
transfer times are exclusively a function of bandwidth. Cache bandwidths are
typically well documented and can be found in vendor data sheets. Depending on
how many \ac{CL}s have to be transferred, the contribution of each level in the
memory hierarchy ($T_\mathrm{L1L2}$, \ldots, $T_\mathrm{L3Mem}$) can be
determined.  Special care has to be taken when dealing with main memory
bandwidth, because peak memory bandwidth specified in the data sheet and
sustained memory bandwidth $b_\mathrm{s}$ can differ greatly. In addition, in
practice the sustained bandwidth may also depend on the number of distinct load and store
streams. It is therefore recommended to empirically determine $b_\mathrm{s}$
using a kernel that resembles the memory access pattern of the benchmark to be
modeled.  Once $b_\mathrm{s}$ has been obtained, the time to transfer one
\ac{CL} between the cache hierarchy and main memory can be derived from the CPU
frequency $f$ as $64\,\mathrm{B} \cdot f / b_\mathrm{s}$~cycles.

In a second step, an empirically determined latency penalty $T_p$ is applied to
off-core transfer times. This departure from the bandwidth-only model has been
mandated by the inability of some architectures to hide the memory access
latency.
On regular
Xeon processors, this penalty is added for each level in the memory hierarchy
that has to make use of the Uncore interconnect (i.e., the L3 cache, as data is
pseudo-randomly distributed between all last-level cache segments and memory,
because the memory controller is attached to the ring bus as well). On Knights
Corner there exists no shared cache when each thread is working on its own
data; each core is using data from its local L2 cache so the latency penalty is
only added when the core-ring-interconnect is used to get data from main
memory.  Instruction times as well as data transfer times, e.g.,
$T_\mathrm{L1L2}$ for the time required to transfer data between L1 and L2
caches, are summarized in a shorthand notation:
\ecmshort{T_\mathrm{OL}}{T_\mathrm{nOL}}{T_\mathrm{L1L2}}{T_\mathrm{L2L3}+T_\mathrm{p}}{T_\mathrm{L3Mem}+T_\mathrm{p}}.

To arrive at a prediction, in-core execution and data transfer times must be combined 
appropriately.
The runtime is given by either $T_\mathrm{OL}$ or the sum of
non-overlapping core cycles $T_\mathrm{nOL}$ plus contributions of data
transfers $T_\mathrm{data}$, whichever takes longer. $T_\mathrm{data}$ 
comprises all necessary data
transfers in the memory hierarchy, plus latency penalties if applicable. 
Again we have to distinguish between overlapping and non-overlapping
behavior; in case of Intel Xeon, any data transfer during a specific cycle 
in the inclusive
cache hierarchy prevents all other transfers (including those in 
$T_\mathrm{OL}$) in that cycle. $T_\mathrm{data}$ is thus the sum of all
cycles required to transfer the data to L1 and back. E.g., 
for data coming from the L3 cache we have
$T_\mathrm{data}=T_\mathrm{L1L2}+T_\mathrm{L2L3}+T_\mathrm{p}$. The prediction
is thus $T_\mathrm{ECM} = \max(T_\mathrm{OL}, T_\mathrm{nOL} +
T_\mathrm{data})$. Note that for other architectures with different overlapping properties 
and/or exclusive cache hierarchies this formula may look very different. 

In order to summarize the predictions for data coming 
from different levels in the hierarchy we use a shorthand notation:
\ecmpshort{T_\mathrm{ECM}^\mathrm{core}}{T_\mathrm{ECM}^\mathrm{L2}}{T_\mathrm{ECM}^\mathrm{L3}}{T_\mathrm{ECM}^\mathrm{Mem}}{}.
Converting from time (cycles) to performance is done by dividing the work $W$
(e.g., floating-point operations, updates, or any other relevant work metric) 
by the runtime: $P_\mathrm{ECM} = W/T_\mathrm{ECM}$.

\begin{figure}[tb]
\centering
\includegraphics*[width=1.0\linewidth]{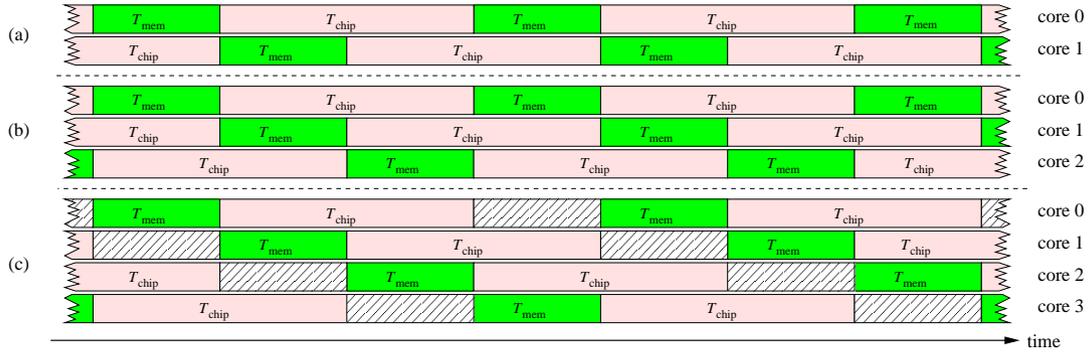}
\caption{\label{fig:ecm_mc_scaling}Multicore scaling in the ECM model. $T_\mathrm{mem}$ 
  is the time spent for data transfers over the memory bottleneck, while $T_\mathrm{chip}$ 
  comprises all non-bottlenecked (i.e., core-local) contributions. The saturation point 
  is at three cores in this case since $(T_\mathrm{chip}+T_\mathrm{mem})/T_\mathrm{mem}=3$\@. 
  Hatched boxes denote stalls, which emerge from 
  using more cores than needed for bandwidth saturation.}
\end{figure}
The model assumes that single-core performance scales linearly with the cores until a shared
bottleneck is saturated. On most modern processors the only shared bottleneck is main
memory bandwidth. 
As shown in Fig.~\ref{fig:ecm_mc_scaling}, the ratio of the overall single-core
execution time and the contribution of the bottleneck determines 
the maximum speedup: as long as the number of cores is smaller than this ratio,
the memory bus is not saturated. In terms of the ECM model, the maximum
speedup is
$\sigma_\mathrm{S} = T_\mathrm{ECM}^\mathrm{Mem}/T_\mathrm{L3Mem}$. Performance at
the saturation point is then $P_\mathrm{ECM}^\mathrm{S} =  f\cdot \sigma_\mathrm{S} \cdot
W_\mathrm{CL}/T_\mathrm{ECM}^\mathrm{Mem} = f \cdot
W_\mathrm{CL}/T_\mathrm{L3Mem}$, where $W_\mathrm{CL}$ is the work per \ac{CL}
and $f$ is the processor clock frequency. This is just another formulation of the
bandwidth-bound part of the Roof{}line model \cite{roofline:2009}. The core count
necessary to saturate the memory bandwidth is $n_\mathrm{S} = \lceil
T_\mathrm{ECM}^\mathrm{Mem}/T_\mathrm{L3Mem} \rceil$. If $n_\mathrm{S}\geq n_\mathrm{chip}$,
i.e., if the required number of cores for saturation exceeds the available
number, the code is scalable.

\section{Experimental testbed}
\label{sec:testbed}

Table~\ref{tab:arch} gives an overview of the relevant architectural details of
the systems used in this paper. The regular Xeon machines (Haswell-EP and
Broadwell-EP) and the POWER8 machine are standard two-socket systems. The Xeon
Phi coprocessor (Knights Corner) is a PCIe card hosted in a standard two-socket
Ivy Bridge-EP system.
\begin{table}[tb]
\renewcommand{\arraystretch}{1.2}\centering
\begin{tabular}{l@{\hspace{5pt}}c@{\hspace{5pt}}c@{\hspace{5pt}}c@{\hspace{5pt}}c}
        \hline
	Microarchitecture               & Haswell-EP        & Broadwell-EP      & Knights Corner        & POWER8    \\
    Shorthand                       & HSW               & BDW               & KNC                   & PWR8      \\
	Chip model                      & E5-2695 v3        & unknown           & 5110P                 & S822LC  \\
	Release date                    & Q3 2014           & pre-release       & Q4 2012               & Q2 2014   \\
        \hline
	Nominal CPU clock               & 2.3\,\GHZ         & 2.1\,\GHZ         & 1.05\,\GHZ            & 2.926\,\GHZ \\
    Cores/threads                   & 14/28             & 22/44             & 60/240                & 10/80     \\
    Max. SIMD width                 & 32\,\BYTE         & 32\,\BYTE         & 64\,\BYTE             & 16\,\BYTE \\
    \# of SIMD registers            & 16                & 16                & 32                    & 64        \\
        \hline
    Instruction \makebox[0pt][l]{throughput per cycle} & &                   &                       &           \\
        LOAD/STORE                  & 2 / 1             & 2 / 1             & 1 / 1                 & 2 / 2       \\
        ADD/MUL/FMA                 & 1 / 2 / 2         & 1 / 2 / 2         & 1 / 1 / 1             & 2 / 2 / 2   \\
        \hline
        Core-private caches         & 32\,kB L1         & 32\,kB L1         & 32\,kB L1             & 64\,kB L1         \\
                                    & 256\,kB L2        & 256\,kB L2        & 512\,kB L2            & 512\,kB L2        \\
                                    & ---               & ---               & ---                   & 8\,MB L3          \\
        Shared caches               & 35\,MB L3         & 55\,MB L3         & ---                   & 64\,MB L4         \\
        L2-L1 bandwidth             & 64\,\BYTE/cy      & 64\,\BYTE/cy      & 32\,\BYTE/cy          & 64\,\BYTE/cy      \\
        L3-L2 bandwidth             & 32\,\BYTE/cy      & 32\,\BYTE/cy      & ---                   & 32\,\BYTE/cy       \\
        MEM-L3 bandwidth            & $\sim$14\,\BYTE/cy& $\sim$15\,\BYTE/cy& 160\,\BYTE/cy         & ---                \\
        Centaur-L2 bandwidth        & ---               & ---               & ---                   & $\sim$19\,\BYTE/cy \\
	    Main memory                 & 4$\times$DDR4-2166   & 4$\times$DDR4-2166   & 16$\times$GDDR5-5000     & 4$\times$Centaur   \\
        Theor. load BW                & 69.3\,GB/s        & 69.3\,GB/s        & 320\,GB/s             & 76.8\,GB/s        \\
        Meas. load BW                & 2$\times$32.0\,\GBS\ (92\%)& 2$\times$32.3\,\GBS\ (93\%)& 175\,\GBS\ (55\%)     & 73.6\,\GBS\ (96\%) \\
        \hline
	\end{tabular}\\[3mm]
    \caption{Test machine specifications and micro-architectural features (one socket). The cache line size is 64\,\bytes\ for all Intel architectures and 128\,\bytes\ for IBM POWER8.}
    \label{tab:arch}
\end{table}

Note that BDW corresponds to a ``tick'' in Intel's design model, i.e.,
it is a shrink in the manufacturing process technology from 22\,nm to
14\,nm with only minor architectural improvements compared to
HSW\@.  All results for Broadwell-EP are preliminary since we
only had access to a pre-release version of the chip.

All SIMD instructions set extensions for the covered microarchitectures
support \ac{FMA} instructions. The vector
scalar extension (VSX) on IBM's PWR8 have a SIMD width of 16\,\BYTES.
The AVX2 vector extensions supported by HSW and BDW have a
SIMD width of 32\,\BYTES\ and KNC's initial many core instructions
(IMCI) allow for 64-\BYTE\ SIMD. All Intel processors employ a fully
inclusive cache architecture whereas PWR8 uses an exclusive victim
cache architecture for the last level cache. This results in different data
paths inside the caches. On PWR8 data is loaded from memory directly into the
L2 caches, and only cachelines which get evicted from L2 will be copied back to the
L3 cache.

The sustained memory bandwidth for all architectures was determined using
a naive dot product benchmark. To obtain good results on the Xeon Phi, we
followed the optimization instructions for the STREAM benchmark as described by
Intel \cite{Intel-STREAM-Phi}; in particular, we set the prefetching distance
64 \ac{CL}s ahead for the L2 cache, 8 \ac{CL}s for the L1 cache, and used one
thread per core to avoid congestion on the ring bus. The IBM PWR8 memory
bandwidth requires further explanation. PWR8 uses a custom high frequency
channel interface between the processor chip and a memory buffer chip
(Centaur) \cite{IBM-mem}. Each Centaur chip connects to four DRAM channels.
PWR8 supports up to eight memory channels per chip operating at 9.6\,\GHZ\
with a bus width of 2\,\BYTE\ (read) plus 1\,\BYTE\ (write)\@. Our test system is an IBM
S822LC and supports only four Centaur chips. The four memory channels can
provide up to 115.2\,\GBS\ read/write or 76.8\,\GBS\ read-only bandwidth per chip. Note
that this is significantly less than what the 16 attached DRAM channels
(DDR3-1333) could provide (170.6\,\GBS). A fully equipped high-end PWR8
system hence has twice the memory bandwidth per chip.

Unless noted otherwise, KNC was used in 2-SMT and PWR8 in 8-SMT
mode, i.e., two respectively eight threads were run on each physical core.  On
HSW and BDW a single thread was run on each physical core, and
Uncore frequency scaling was deactivated. Furthermore, the ``cluster on die'' (COD) mode
was active for HSW and BDW.  In CoD mode, the chip is logically split
into two ccNUMA domains of equal size. Last-level cache and
memory requests are limited to the domain a core is assigned to, reducing
latency and collisions in the Uncore interconnect. The two
memory domains per chip are visible in the load-only bandwidth row of
Table~\ref{tab:arch}; e.g., the sustained load-only bandwidth for HSW is
32.0\,\GBS\ per memory domain and 64.0\,\GBS\ per chip. For details on
the CoD mode see \cite{HFEHW15a}.

\section{Optimal implementations and performance models for dot}
\label{sec:impl}

We only discuss variants for dot in \ac{SP} here.
The model prediction in terms of cycles per \ac{CL} does not change for the
SIMD variants of Kahan when going from \ac{SP} to \ac{DP}, but one \ac{CL} update
represents twice as much useful work (scalar iterations) in the \ac{SP}
case.
To eliminate variations introduced by compiler-generated code we implemented all kernels directly
in assembly language and use the \verb.likwid-bench. microbenchmarking framework~\cite{Treibig:2011:3} to perform measurements.

\subsection{Naive scalar product}
An optimal implementation of the naive scalar product in single precision serves as the baseline (see
Fig.~\ref{fig:listings}a). All versions of the Kahan-enhanced scalar product
described in Section~\ref{sec:kahan} will be compared to this baseline.
\begin{figure}[tb]
(a)\begin{minipage}[t]{0.43\linewidth}
\begin{lstlisting}
  float sum = 0.0;

  for (int i=0; i<N; i++) {
     sum = sum + a[i] * b[i]
  }
  \end{lstlisting}
\end{minipage}\hfill
(b)\begin{minipage}[t]{0.46\linewidth}
\begin{lstlisting}
    float sum = 0.0;
    float c = 0.0;
    for (int i=0; i<N; ++i) {
        float prod = a[i]*b[i];
        float y = prod-c;
        float t = sum+y;
        c = (t-sum)-y;
        sum = t;
    }
\end{lstlisting}
\end{minipage}
  \caption{\label{fig:listings}(a) Naive scalar product code in single precision.
    (b) Kahan-compensated scalar product code.}
\end{figure}

Sufficient unrolling must be applied to hide the ADD pipeline latency for the
recursive update on the accumulation register and to apply SIMD vectorization.
Both optimizations introduce partial sums and are therefore not compatible with
the C standard as the order of non-associative operations is changed. With
higher optimization levels (\texttt{-O3}) the current Intel C compiler (version
15.0.2) and IBM XL C compiler (version 13.1.3) both generate optimal
code. Note that partial sums usually improve the accuracy of the result
\cite{Dalton:2014}.

\subsubsection{Intel Haswell-EP and Broadwell-EP}

On HSW and BDW the kernel is limited by the throughput of the
LOAD units (see Table~\ref{tab:arch}).  Two AVX loads per vector (\verb.a. and
\verb.b.) are required to cover one unit of work (16 scalar loop iterations),
leading to a total of four AVX load instructions; with two LOAD units, the core
can execute two LOAD instructions per cycle, resulting in
$T_\mathrm{nOL}=2\,\cycles$. To process the data, two \ac{FMA} instructions have
to be executed; with two \ac{FMA} units, the core can execute both
instructions in a single cycle, resulting in an overlapping part of
$T_\mathrm{OL}=1\,\cycles$.

If the data is in the L2 cache, two \ac{CL}s (one each for \texttt{a} and \texttt{b})
have to be transferred to the L1 cache; at the advertised bandwidth
of 64\,\BYTE/\cycle\ this results in $T_\mathrm{L1L2}=2\,\cycles$. With data in L3
it takes $T_\mathrm{L2L3}=4\,\cycles$ to transfer the
two \ac{CL}s to L2 due to the L2-L3 bandwidth of 32\,\BYTE/\cycle; the empirical latency
penalty was determined to be $T_\mathrm{p}=1\,\cycles$ for the 14-core
HSW and $T_\mathrm{p}=5\,\cycles$ for the 22-core BDW\@. 
The latency penalty is strongly correlated with the number of hops in the Uncore; as
BDW features more cores and each core's L3 slice forms a hop in the
Uncore its latency is higher than that of the HSW chip with fewer
cores/hops.

To compute the contribution of transferring the two \ac{CL}s from main memory
to the L3 cache, we convert the sustained memory bandwidth from \GBS\ to \BYTE/\cycles.
Note that in cluster on die mode a single core can only make use of the
bandwidth inside its memory domain. For the HSW, which runs at 
2.3\,\GHZ, the measured memory domain bandwidth of 32.0\,\GBS\
corresponds to a transfer time of
$64\,\mathrm{\BYTE/CL}\cdot2.3\,\GHZ/32.0\,\GBS=4.6\,\cycles/\mathrm{CL}$
or $9.2\,\cycles$ for both \ac{CL}s. BDW runs at
2.1\,\GHZ, which leads to a transfer time of
$64\,\mathrm{\BYTE/CL}\cdot2.1\,\GHZ/32.3\,\GBS=4.2\,\cycles/\mathrm{CL}$
or $8.4\,\cycles$ for both \ac{CL}s. The same latency penalty as for the L3 cache is
applied for data coming from main memory, because the data has to be moved from
the memory controller to the L3 cache segment in which the cache line is
placed. The resulting ECM model inputs are \ecm{1}{2}{2}{4+1}{9.2+1}{\cycles} for
HSW and \ecm{1}{2}{2}{4+5}{8.4+5}{\cycles} for BDW.

The full ECM prediction reads \ecmp{2}{4}{9}{19.2}{\cycles} for HSW.
We choose an ``\updateX'' (two\,\flops) as the basic unit of work to make
performance results for different implementations comparable. The resulting
unit is ``updates per second'' ($\US$)\@.  The
expected single core performance for the HSW is thus
\bq
P=\frac{16\,\updates\cdot 2.3\,\GCS}{\ecmp{2}{4}{9}{19.2}{\cycles}} =
\ecmp{18.40}{9.20}{4.09}{1.92}{\GUS}\eos
\eq
The predicted saturation point is at $n_\mathrm S=\left\lceil 19.2/9.2\right\rceil=3$
cores per memory domain or 6 cores per chip. Performance at the saturation point is
$P_\mathrm{ECM}^\mathrm{S} = f \cdot W_\mathrm{CL}/T_\mathrm{L3Mem}=2.3\GHZ \cdot 16\,\mathrm{updates}/9.2\,\cycles=4\,\GUS$ per memory domain or 8\,\GUS\ per chip.

For BDW the full ECM prediction is \ecmp{2}{4}{13}{26.4}{\cycles}
and the resulting expected serial performance at 2.1\,\GHZ\ is
\bq
P=\frac{16\,\updates\cdot 2.1\,\GCS}{\ecmp{2}{4}{13}{26.4}{\cycles}} =
\ecmp{16.80}{8.40}{2.58}{1.27}{\GUS}\eos
\eq
The predicted saturation point is at
$n_\mathrm S=\left\lceil 26.4/8.4\right\rceil=4$ cores per memory domain or 8
cores per chip. The difference in sustained memory bandwidth between our
HSW and BDW systems are marginal, so the prediction for the saturated performance
is identical to that of the HSW machine.

\subsubsection{Intel Xeon Phi}

KNC's \ac{IMCI} extensions have a SIMD width of 512\,\bit\ or 64\,\BYTE,
corresponding to a full cache line.  This means that two 512-\bit\ IMCI load
instructions are needed to load the data from the L1 cache into registers, 
so $T_\mathrm{OL}=2\,\cycles$. Processing the data
requires one \ac{FMA} instruction, which has a maximum throughput of one per cycle,
resulting in $T_\mathrm{OL}=1\,\cycles$.  Note that while a KNC
core is much simpler than its HSW or BDW counterpart, it is still
capable of retiring two instructions in a superscalar fashion. It
features two pipelines: a vector pipeline (U-pipe) with the 512-\bit\
vector processing unit attached and a scalar pipeline that handles all
remaining instructions. While SIMD vector arithmetic is only possible on the
U-pipe, the V-pipe can also be used for SIMD load instructions. It is thus
possible to overlap the \ac{FMA} that is scheduled on the U-pipe with one of the
load instructions when both instructions are paired.\footnote{Instruction
pairing happens whenever an instruction scheduled for the U-pipe is followed
directly by an instruction scheduled on the V-pipe.  Restrictions about when
instructions pairing can happen are complex but well documented
\cite{Intel-Pairing}.}

At a bandwidth of 32\,\BYTE/\cycles\ \cite{Intel-Pairing}, it takes
$T_\mathrm{L1L2}=4\,\cycles$ to deliver the data from the L2 cache to the core.
At a clock speed of 1.05\,\GHZ\ and a sustained
memory bandwidth of 175\,\GBS\ the transfer time of a single \ac{CL} is
$64\,\mathrm{\BYTE/CL}\cdot1.05\,\GHZ/175\,\GBS=0.4\,\cycles$, thus
0.8\,\cycles\ for both \ac{CL}s. The empirically
determined latency penalty for the ring interconnect amounts to
$T_\mathrm{p}=20\,\cycles$. The resulting ECM input is
\ecmknc{1}{2}{4}{0.8+20}{\cycles}.

The full ECM prediction is
\ecmpknc{2}{6}{26.8}{\cycles}. It is clear from these numbers that the KNC
is a strongly latency-dominated machine beyond the L2 cache. 
The expected performance of a single core is
\bq
P=\frac{16\,\updates\cdot 1.05\,\GCS}{\ecmpknc{2}{6}{26.8}{\cycles}} =
\ecmpknc{8.40}{2.80}{0.63}{\GUS}\eos
\eq
The predicted saturation point is at
$n_\mathrm S=\left\lceil 26.8/0.8\right\rceil=34$ 
cores, and the maximum performance is $21.3\,\GUS$. 

\subsubsection{IBM POWER8}

In contrast to Intel Xeon and Xeon Phi chips, where cache lines are
64\,\BYTE, the cache line size on the IBM PWR8 is 128\,\BYTE. At a SIMD width of
16\,B this means that a total of 16 VSX LOAD instructions
are required to move data from the L1 cache to the registers,
which takes eight cycles. The L1 cache is multi-ported, i.e., it can supply
data to the registers and simultaneously receive data from the L2 cache
\cite{IBM-core-microarch}. Eight VSX FMA instructions process the data from both
\ac{CL}s; the kernel is thus limited by the
throughput of the LOAD units and $T_\mathrm{OL}=8\,\cycles$. As there are no
non-overlapping instructions, we have $T_\mathrm{nOL}=0\,\cycles$.

Data can be delivered from the L2 to the L1 at a bandwidth of 64\,\BYTE/\cycles, 
thus
$T_\mathrm{L1L2}=4\,\cycles$. Using the documented L2-L3 bandwidth of
32\,\BYTE/\cycles\ we calculate $T_\mathrm{L2L3}=8\,\cycles$. When data is 
in main memory, the bandwidth of the chip-to-Centaur interconnect proves to be the
bottleneck: each centaur can provide 19.2\,\GBS, which translates
into a peak bandwidth of 76.8\,\GBS\ for our system. 
The measured sustained memory bandwidth is 73.6\,\GBS, hence a \ac{CL} transfer takes
$128\,\mathrm{\BYTE/CL}\cdot2.9\,\GHZ/73.6\,\GBS=5.0\,\cycles$.
Consequently, $T_\mathrm{L2L4}=10\,\cycles$.


The resulting ECM input is
\ecm{8}{0}{4}{8}{10}{\cycles}. We assume a latency penalty $T_\mathrm{p}$ of
zero, because in the measurements there is no deviation from the model
prediction for data coming from the L3 cache. The reason is that on PWR8,
each core has a dedicated L3 cache in which data for a particular core
resides; opposed to Intel's Uncore design, no transfers across the L3 cache
interconnect are necessary when accessing data from L3.
The full ECM model prediction is \ecmp{8}{8}{12}{22}{\cycles}.
The predicted saturation point is at
$n_\mathrm S=\left\lceil 22/10\right\rceil=3$ cores.

\subsection{Kahan-enhanced scalar product}
\label{sec:kahan}

Figure~\ref{fig:listings}b shows the implementation of the Kahan algorithm for
the dot product. Compilers have problems with this loop code for two reasons: First, the
compiler detects (correctly) a loop-carried dependency on \verb.c., which prohibits
SIMD vectorization and modulo unrolling. Second, the compiler may recognize
that, arithmetically, \verb.c. is always equal to zero. With high optimization
levels it may thus reduce the code to the naive scalar product, defeating the
purpose of the Kahan algorithm. This is the reason why we use hand-coded
assembly throughout this work. For comparison we also show compiler-generated
Kahan code for which we ensured (by appropriate compiler options)
that the algorithm is preserved.

\subsubsection{Intel Haswell-EP and Broadwell-EP}
One iteration comprises one multiplication, four additions or subtractions, and
two loads. The bottleneck on the HSW and BDW cores is thus the ADD unit
(ADD and SUB are handled by the same pipeline).
In the following we construct the \ecmm\ for the AVX versions of the
Kahan loop.

%
%
%
In an AVX vectorized version of the Kahan-enhanced dot product kernel 
that does not use the new FMA3 extensions, we
require two AVX multiplication instructions and eight AVX
additions/subtractions to process one unit of work (eight scalar iterations)\@.
Multiplications can be executed speculatively several loop iterations ahead,
because they have no data dependencies. This means that the five (HSW)
respectively three (BDW) cycles of latency of the multiplication are not
an issue. With at least four-way unrolling the add latency of three cycles can be
hidden. The throughput is thus limited by the ADD unit, on which both AVX
additions and subtractions are executed, resulting in
$T_\mathrm{OL}=8\,\cycles$\@.
Because data movement is exactly the same as in the naive dot product, the
remaining model inputs stay the same.  This results in the following inputs for
the ECM model:
\ecm{8}{2}{2}{4+1}{9.2+1}{\cycles} for HSW and
\ecm{8}{2}{2}{4+5}{8.8+5}{\cycles} for BDW. The resulting ECM
predictions are \ecmp{8}{8}{9}{19.2}{\cycles} and
\ecmp{8}{8}{13}{26.8}{\cycles}, respectively.

\begin{SCfigure}[0.9][tb]
    \includegraphics*[width=0.75\linewidth]{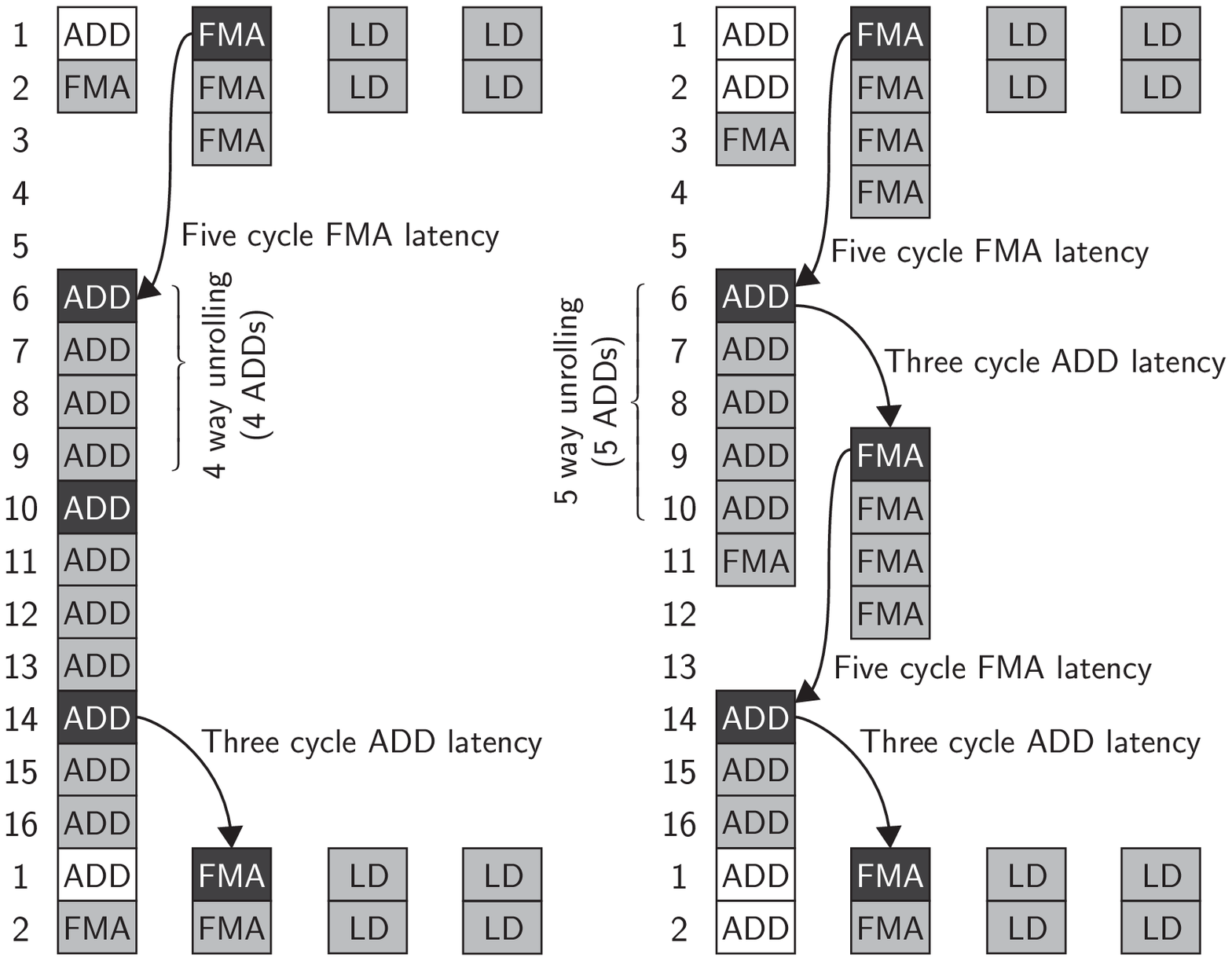}
    \caption{\label{fig:FMA3-sched}Left: Four-way unrolled Kahan dot product kernel using FMAs. 
      Right: Our five-way unrolled optimized version.}
\end{SCfigure}

At first glance, when making use of the new \ac{FMA} instructions we
expect the number of in-core cycles to drop, because each core can execute two
\ac{AVX} \ac{FMA} instructions per cycle. The multiplication in line four and
the subtraction in line five of the source code in Fig.~\ref{fig:listings}b
can be handled by a single \texttt{vfmsub231ps}
instruction. This reduces the number of additions/subtractions to six per cache
line update so we expect $T_\mathrm{OL}$ to drop to six cycles. However, the
situation is more complicated. Since the \ac{FMA} instructions have
\texttt{y} as input, the instruction can no longer be executed speculatively, which
means that the ADD instructions now have to wait for the \ac{FMA}
instruction, which has a five-cycle latency on both HSW and BDW.
Unfortunately, 16 addressable AVX registers are not enough to
perform sufficient unrolling to completely hide this latency. It turns
out that a four-way unrolled loop results in the same
$T_\mathrm{OL}$ of eight cycles (see left part of
Fig.~\ref{fig:FMA3-sched}). For a four-way unrolled kernel, intra-loop
latencies play a significant role: After the first \ac{FMA} has been scheduled in
the first cycle of the loop (shown in black), it takes five cycles until the
addition using the result of the \ac{FMA} can be issued in cycle six (shown in
black). The three-cycle latency of the addition is hidden by using four-way
unrolling; thus it takes four cycles until the next addition corresponding to
the partial sum can be retired. Finally, in the fourteenth cycle, the last
addition of the first partial sum is issued. Only after the ADD latency of
three cycles, the first \ac{FMA} of the next loop iteration can be issued, because
it uses the result of the addition as input.

Even by reusing registers that can be overwritten because their content is no
longer needed, the maximum unrolling factor that can be achieved is five.
Unrolling the loop with the previous strategy will result in an execution time
of 18 cycles\footnote{By increasing the unrolling factor from four to five, we
have to wait $2\times5$ cycles after the ADD instructions instead of $2\times4$
when using four-way unrolling.} for one loop iteration (handling 2.5 cache
lines at 5-way unrolling) corresponding to 7.2\,\cycles/CL. It is possible to further
decrease the runtime by ``abusing'' \ac{FMA} operations: by keeping a
vector register that has all its components set to floating-point one, we can
model an addition, i.e.,  $y=a\times1.0+b$\@. By
this optimization we can keep the loop iteration time at 16 cycles for 5-way
unrolling, corresponding to a $T_\mathrm{OL}=6.4\,\cycles$. The instruction
scheduling for this version is shown on the right in
Fig.~\ref{fig:FMA3-sched}. We replace the second unrolled additions by an \ac{FMA}
to increase throughput while minimizing the five-cycle latency
via unrolling. The ECM model input for this optimized kernel is
\ecm{6.4}{2}{2}{4+1}{9.2+1}{\cycles} for HSW and
\ecm{6.4}{2}{2}{4+5}{8.8+5}{\cycles} for BDW. The resulting ECM
predictions are \ecmp{6.4}{6.4}{9}{19.2}{\cycles} and
\ecmp{6.4}{6.4}{13}{26.8}{\cycles}, respectively.

The conclusion from this analysis is that there is no expected performance
difference for in-memory working sets between the naive scalar product and the
Kahan version if AVX vectorization is applied to Kahan. It comes for free even
in the L3 cache. Only for in-L1 and in-L2  data we expect a 2$\times$ slowdown
for Kahan versus the naive version even with the best possible code.

\subsubsection{Intel Xeon Phi}

\begin{figure}[tb]
\lstset{
        breaklines=true,
        language=C,
        basicstyle=\small\ttfamily,
        numbers=left,
        numberstyle=\tiny,
        frame=t,
        columns=fullflexible,
        showstringspaces=false,
        escapechar=\%,
        escapebegin=\color{blue}\ttfamily\bfseries,
        keepspaces=true
}
\begin{lstlisting}
    vfmsub231ps zmm2, zmm0, [rdx+rax*8] # y=A[i]*B[i]-c
    vprefetch0 [576+rsi+rax*8]          # prefetch A into L1

    vaddps zmm4, zmm3, zmm2             # t=sum+y
    vmovaps zmm0, [rsi+rax*8+64]        # load A for next iter

    vsubps zmm5, zmm4, zmm3             # tmp=t-sum
    vprefetch0 [512+rdx+rax*8]          # prefetch B into L1

    vsubps zmm2, zmm5, zmm2             # c=tmp-y
    vmovaps zmm3, zmm4                  # sum=t
\end{lstlisting}

  \caption{\label{fig:knc_asm}Assembly code of the loop body for L2-optimized Kahan-enhanced dot product on KNC.}
\end{figure}

On KNC, the vector instructions performing arithmetic operations can
only retire on the vector U-pipe. Thus it makes no sense to use a similar
strategy as on HSW and BDW to replace additions/subtractions
by fused multiply-add instructions. To process one work unit (16 scalar iterations)
using 512-\bit\ SIMD instructions, the core has to execute one fused
multiply-add and three additions/subtractions, yielding
$T_\mathrm{OL}=4\,\cycles$; the two 512-\bit\ loads can be executed in
parallel with some of the arithmetic instructions when instructions are paired
correctly, resulting in $T_\mathrm{nOL}=2\,\cycles$. At 32\,\BYTE/\cycle,
$T_\mathrm{L1L2}=4\,\cycles$ for two cache lines. As previously determined,
the sustained memory bandwidth of 175\,\GBS\ corresponds to a transfer time of
0.4\,\cycles/CL; thus $T_\mathrm{L2Mem}=0.8\,\cycles$.

We found that it is necessary to use separate, specifically designed kernels
to obtain the best performance for each individual cache level. 
The L1-optimized kernel needs no prefetching instructions at all. 
For data in the L2 cache, two
software prefetching instructions are used, fetching eight cache lines ahead.
These two
instructions can be paired with arithmetic instructions and thus do not
change in-core execution time (see lines two and eight in
Fig.~\ref{fig:knc_asm})\@. For data coming from main memory, we prefetch 
64 iterations ahead into the L2 cache 
and also keep the previous prefetching strategy of fetching cache lines eight
iterations ahead from L2 into L1\@. The two new prefetch instructions can no
longer be paired, because we run out of unpaired arithmetic instructions: The
first \ac{FMA} and the first ADD is paired with the LOADs that bring data
into the registers; the second and third ADD/SUB are paired with
the L2-L1 software prefetch instructions. The in-core
execution time is thus extended by two additional cycles for the two prefetch
instructions from main memory into L2\@.  The ECM input
for KNC thus is
\ecmknc{4}{2+2_\mathrm{L2}+2_\mathrm{MEM}}{4}{0.8+17}{\cycles}. Note that
the composition of $T_\mathrm{nOL}$ is dependent on where input data is coming
from: in the L1 kernel, we are retiring just two load instructions so
$T_\mathrm{nOL}$=2\,\cycles; in the kernel optimized for data coming from the L2
cache, we need to include two prefetching instructions so
$T_\mathrm{nOL}$=2\,\cycles+2\,\cycles=4\,\cycles; finally, for the memory-optimized kernel, we
have to include two more prefetching instructions, so $T_\mathrm{nOL}$=6\,\cycles.
The full ECM prediction is \ecmpknc{4}{8}{27.8}{\cycles}.

\subsubsection{IBM POWER8}

On PWR8, 16 VSX LOADs (eight 16-byte LOADs per 128-byte cache line) and required to
load and an additional 32 (eight VSX FMA and 24 VSX ADD/SUB) instructions are
required to process one cache line. Core throughput is limited by the two
arithmetic VSX units, which require 16 cycles to process all 32 FMA/ADD/SUB
instructions, resulting in $T_\mathrm{OL}=16\,\cycles$; $T_\mathrm{nOL}$ is
zero. The remaining ECM inputs are identical to the naive dot product, yielding
\ecm{16}{0}{4}{8}{10}{} as ECM input. The full ECM prediction is
\ecmp{16}{16}{16}{22}{\cycles}.

\section{Performance results and model validation}\label{sec:results}

\subsection{Intel Haswell-EP and Broadwell-EP}
\begin{figure}[tb]
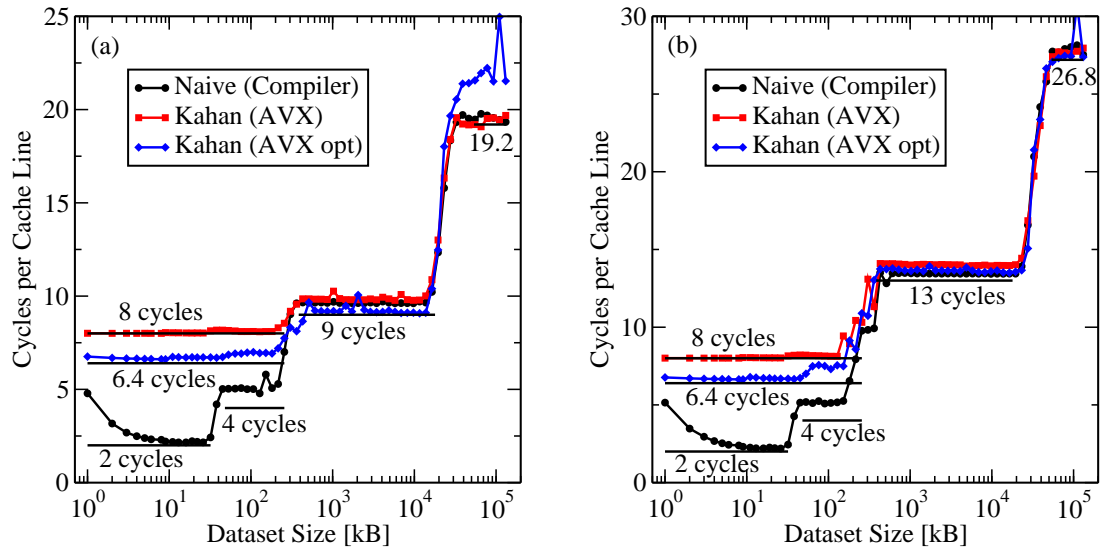

\centering
\includegraphics*[width=0.47\linewidth]{HSW-single-core-SP}\hfill
\includegraphics*[width=0.47\linewidth]{BDW-single-core-SP}
\caption{\label{fig:kahan_hsw_bdw_core} Single-core cycles per CL vs.\
  data set size for AVX, AVX/FMA and the naive scalar product in SP on
  (a) HSW and (b) BDW. The horizontal black lines represent the \ecmm\
  predictions.}
\end{figure}
Single-core benchmarking results for single precision
on HSW and BDW are shown in 
Figs.~\ref{fig:kahan_hsw_bdw_core}a and \ref{fig:kahan_hsw_bdw_core}b. The
model describes the overall behavior very well. The naive (plain sdot) and the AVX Kahan version
show identical performance in L3 cache and beyond. As predicted there is no
performance drop for the AVX Kahan version from L1 to L2. The naive version as well as
the AVX/FMA variant of Kahan fall short of the L2 model prediction; whether this is
due to inefficiencies of the hardware prefetcher or issues with the new 64-\BYTE\
wide bus between L2 and L1 can only be speculated upon. We have no
explanation for why the AVX/FMA optimized version shows worse in-memory
performance on HSW\@.

In-memory scaling results on the chip level for HSW and BDW are shown
in Figs.~\ref{fig:kahan_all_scaling}a and
\ref{fig:kahan_all_scaling}b. Note here that due to the cluster on die mode,
the actual number of cores per memory domain is half of what the $x$ axis in 
the graphs shows, i.e., the two-core run was done with one core per memory
domain. This ensures that we can report the capabilities of the full chip.
The number of cores required to reach saturation is underestimated in
both cases. It is a well-known deficiency of the ECM model that the
scaling behavior near the saturation point is not tracked
correctly. We attribute this to the documented change in the
prefetching strategy near memory bandwidth saturation~\cite{intel-orm-2015}\@.
The
compiler-generated Kahan code is so slow that it misses the target of memory
bandwidth saturation by far on both architectures. On HSW one would need more than
twice the number of available cores to reach saturation.

\subsection{Intel Xeon Phi}

\begin{SCfigure}[0.9][tb]
\centering
\includegraphics*[width=0.5\linewidth]{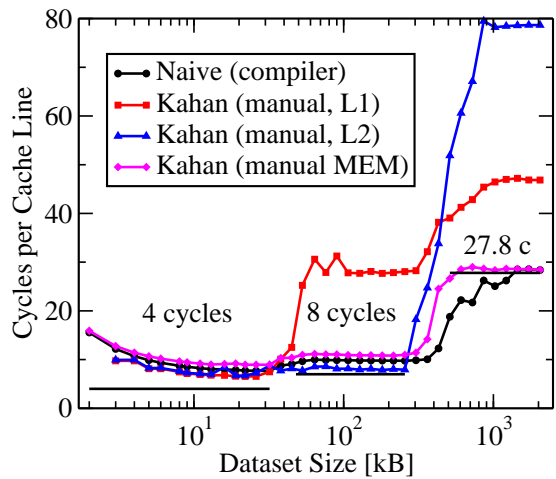}
\caption{\label{fig:kahan_knc_core} Single-core cycles per CL vs.\
  data set size with different implementations tuned for specific memory hierarchy levels of the Kahan scalar
  product and the compiler generated naive scalar product in SP on
  Intel KNC. The black horizontal lines represent the \ecmm\
  predictions. All versions use 2-SMT except the manual memory-optimized kernel, which uses 4-SMT.}
\end{SCfigure}
Figure \ref{fig:kahan_knc_core} shows the \ac{SP} single core results for Xeon Phi.
The model fits very well as long as the special code variant in every memory
hierarchy level is used. Although the Xeon Phi has a hardware prefetcher,
best performance can only be achieved by appropriate software
prefetching.

In-memory scaling results are shown in Fig.~\ref{fig:kahan_all_scaling}c.  In
accordance with Intel's guidelines, which recommend using a single thread per
core when trying to reach the maximum sustained bandwidth on KNC
\cite{Intel-STREAM-Phi}, all in-memory scaling measurements were performed with
1-SMT.  The compiler-generated naive and manual Kahan variants are all but
identical. Xeon Phi exposes a piecewise linear scaling behavior which is not
captured by the linear scaling assumption of the ECM model: Three phases can be
identified, with a clear change in slope at about 20 and 50 cores. While the
naive and manual Kahan codes achieve bandwidth saturation, the naive compiler
version misses it by far.

\subsection{IBM POWER8}\label{sec:PWR8-results}
\begin{figure}[tb]
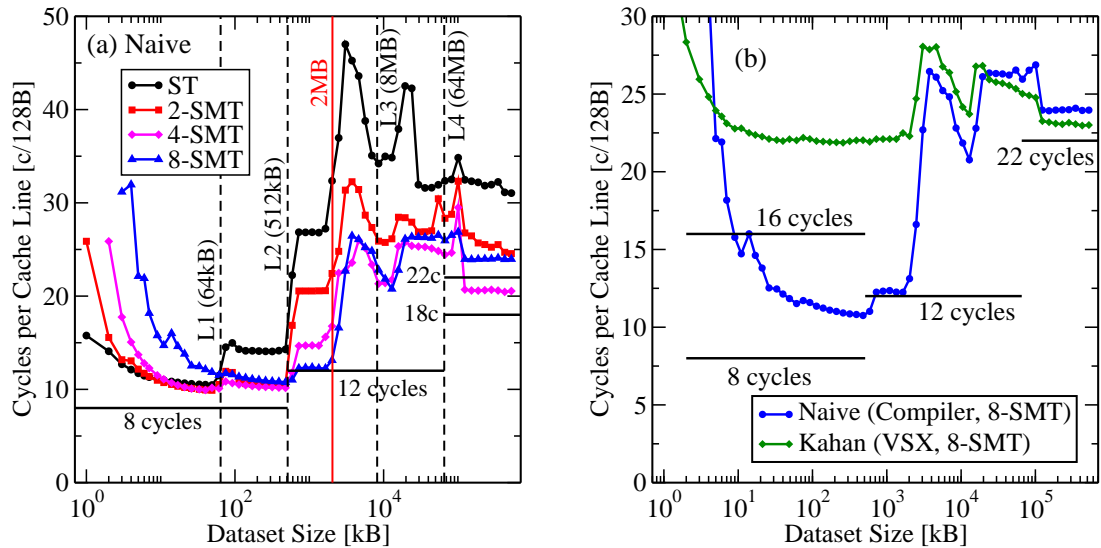

    \includegraphics*[width=0.47\linewidth]{P8-naive-detail}\hfill
    \includegraphics*[width=0.47\linewidth]{P8-single-core-SP}
\caption{\label{fig:kahan_p8_core}
Single-core cycles per CL vs.\ data set size on PWR8. (a) Results for different SMT settings for naive scalar product using SP. (b) Comparison of compiler-generated naive scalar product and manual SIMD Kahan enhanced scalar product using SMT-8.
  The horizontal lines are \ecmm\
  predictions.}
\end{figure}
Fig~\ref{fig:kahan_p8_core}b shows the \ac{SP} single core results for the
PWR8 processor. The model correctly predicts the observed identical
performance in L1 and L2 for the naive variant and in all memory hierarchy levels for the
Kahan variant. In contrast to the Intel architectures 
we failed to reach the predicted 
instruction throughput of the processor by 20--30\%. PWR8 is also 
more sensitive to small loop lengths. The 8\,\MB\ L3 cache is only effective up to 2\,\MB. 
Beyond this point performance dramatically decreases and fluctuates.
The aggregated L4 buffer cache is not visible in the measurements.
For in-memory data sets the performance improves and stabilizes. 
There is no documented hardware feature that could explain the 
erratic behavior between 2\,\MB\ and 64\,\MB\ working set size.

Fig~\ref{fig:kahan_p8_core}a shows the impact of different SMT options on
the naive sdot performance. There is no SMT setting that shows 
competitive performance in all memory hierarchy levels. 
In L1, more SMT threads lead to shorter loops and a corresponding breakdown in 
performance. 
In L2, any number of threads greater than one enables ``wirespeed.''
In L3 (up to 2\,\MB) there is clearly is a strong latency effect, 
which can be compensated only by SMT-8\@. 
From 2\,\MB\ to the L4 capacity limit all variants exhibit
the same fluctuating performance pattern with SMT-4 and SMT-8 showing the best performance. Then in
memory surprisingly SMT-4 is significantly better than SMT-8. For in-memory data sets we
provide two \ecmm\ predictions: 18\,\cycles\ if we assume that evicts of
cachelines from L2 to L3 fully overlap with reloads from memory to L2, and
22\,\cycles\ if we assume there is no overlap among those contributions. Only
SMT-4 is faster than 22\,\cycles, indicating that there is at least 
some overlap. More investigations are necessary to fully understand this
complex behavior.
\begin{figure}[tb]
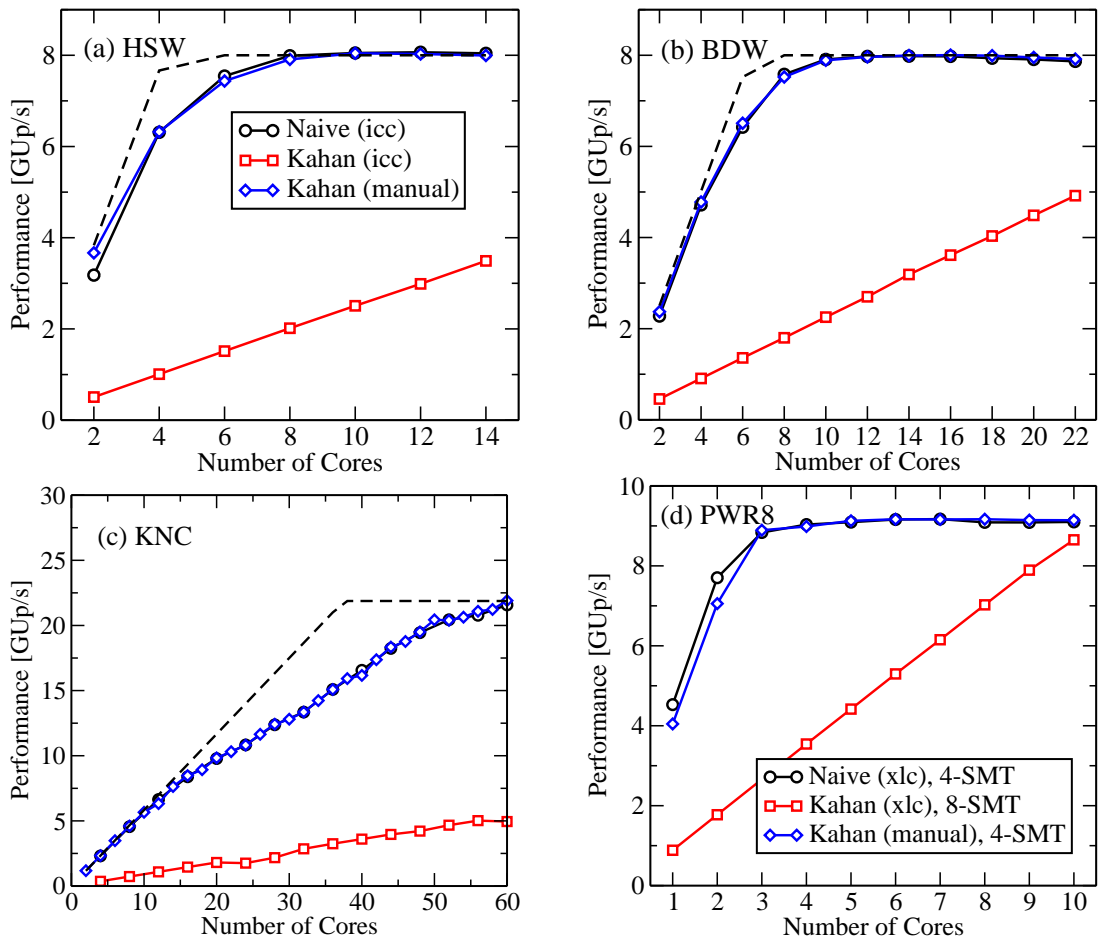

    \includegraphics*[width=0.47\linewidth]{HSW-core-scaling-SP}\hfill
    \includegraphics*[width=0.47\linewidth]{BDW-core-scaling-SP}\\
    \includegraphics*[width=0.47\linewidth]{KNC-core-scaling-SP}\hfill
    \includegraphics*[width=0.47\linewidth]{P8-core-scaling-SP}\\
    \caption{\label{fig:kahan_all_scaling} In-memory scaling (10\,\GB\ working
        set size) for different implementations of the Kahan scalar product
        using SP on (a) HSW, (b) BDW, (c) KNC,
        and (d) PWR8.  One update (\UP) is equivalent to five \flops\ (one
    MULT, four ADDs).}
\end{figure}

In-memory scaling results are shown in Fig.~\ref{fig:kahan_all_scaling}d.
The Naive and Kahan variants show almost identical scaling behavior and quickly
saturate the memory bandwidth. In contrast to the Intel architectures the
compiler version of Kahan (using SMT-8) almost saturates the bandwidth.

\subsection{DP performance for compiler-generated Kahan variant}
\begin{SCfigure}[0.9][tb]
    \includegraphics*[width=0.47\linewidth]{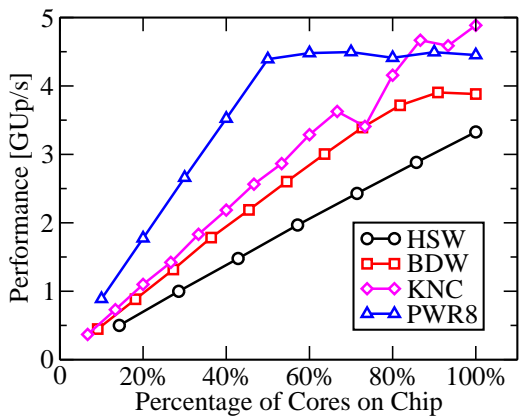}
    \caption{\label{fig:ALL-core-scaling-DP} On-chip performance
      scaling of the compiler-generated Kahan-enhanced ddot 
      on all tested processors. The saturated performance is
      4\,\GUS\ for HSW/BDW, 10.6\,\GUS\ for KNC, and 4.5\,\GUS\ for
      PWR8.}
\end{SCfigure}
As most applications rely on compiler-generated code, we show
the saturation behavior of the compiler-generated Kahan variant for 
\ac{DP} in Fig.~\ref{fig:ALL-core-scaling-DP}\@. Since all compilers
fail at SIMD vectorization, it is interesting to see on which architectures
memory bandwidth is still achieved. On PWR8 we have already observed
near-saturation in the \ac{SP} case; with \ac{DP} this happens at 
five cores. Comparing HSW and BDW, the additional cores help 
BDW to just about saturate whereas HSW misses
this goal. KNC, as expected, misses saturation by a long shot but still achieves
an absolute performance slightly better than PWR8.

\subsection{Comparison across architectures}
\begin{figure}[tb]
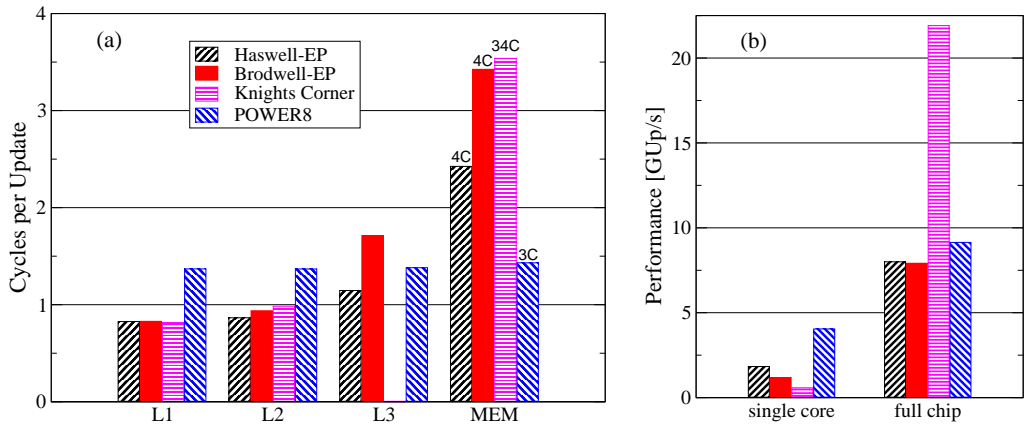

\hspace*{\fill}\includegraphics*[width=0.55\linewidth]{single-core-SP-arch-bar}\hfill
\includegraphics*[width=0.35\linewidth]{arch-bar-GUPs}\hspace*{\fill}
\caption{\label{fig:kahan_single_core_archs}Comparison between
  all tested architectures using the manually implemented SIMD 
  variants of the Kahan-enhanced scalar product in SP:
  (a) Measured single-core runtime in cycles per update in
  different memory hierarchy levels. The saturation point $n_\mathrm S$ is
  indicated above the bars for the memory-bound case (smaller is better). (b)
  Measured full chip performance for the in memory data set (bigger is better).}
\end{figure}
For meaningful cross-architectural comparison of the Kahan-enhanced dot product 
performance we report the cycles \emph{per update} in all memory hierarchy levels
(Fig.~\ref{fig:kahan_single_core_archs}a) and the absolute performance for the
in-memory case in \GUS\ for single core as well as the full chip
(Fig.~\ref{fig:kahan_single_core_archs}b). In L1 and L2 all Intel
architectures run close to their design specifications. PWR8 in
contrast is slightly less efficient missing its design instruction throughput by 30\%.
In L3 and memory the results are
reversed, here the Intel architectures show a significant drop in performance
for L3 and also memory, especially BDW with its complex Uncore design and large
number of cores, whereas PWR8 due to its lock-free memory hierarchy shows less
severe performance breakdowns with increasing working set size.
(Note, however, the large performance variations in a data set size window 
between 2\,\MB\ to over 64\,\MB\ as described in Sect.~\ref{sec:PWR8-results}.)

Regarding absolute single-core and full-chip in-memory performance
(Fig.~\ref{fig:kahan_single_core_archs}b), PWR8 due to its cache
architecture and higher frequency shows the best performance of all
multicore chips, only surpassed by the full-chip KNC by more than a factor
of two due to the latter's superior memory bandwidth.
\section{Conclusion}\label{sec:conc}
We have investigated the performance of naive and Kahan-enhanced
variants of the scalar product on a range of recent multi- and manycore chips.
Using the \ecmm\ the single-core performance in all
memory hierarchy levels and the multi-core scaling for in-memory data
were accurately described. The most important result is that even the
single-threaded optimized Kahan implementation comes with no performance penalty 
on the
Intel multicore chips under investigation
compared
to a naive sdot implementation  in the
L3 cache and in memory. On IBM POWER8 this applies
only for in-memory data sets. 
On the other hand, the POWER8 is able to saturate the memory bandwidth 
with very few cores and provides the best single-core and chip-level
performance for in-memory data. 
Depending on the particular architecture and whether single or double precision
is used, even compiler-generated code may achieve memory bandwidth saturation
on the full chip. Intel Xeon Phi as well as IBM POWER8 require special code or SMT
settings to achieve best performance in different memory hierarchy levels.
Further investigations are necessary to explain erratic performance
behavior on POWER8 for data sets between 2\,\MB\ and 64\,\MB.

We emphasize that the approach and insights described here for the special
case of the Kahan scalar product can serve as a blueprint for other
load-dominated streaming kernels. Especially on POWER8, the \ecmm\ still 
needs to be validated and adjusted 
using more complex codes such as stencil algorithms.

\subsubsection{Acknowledgement}

We thank pro com and IBM Germany for access to an IBM POWER8 test system, and Intel Germany for providing an early access
Broadwell-EP test system.

\bibliographystyle{wileyj}
\bibliography{publications}
\end{document}